\documentclass[
 reprint,
 amsmath,amssymb,
 superscriptaddress,
 aps,
 prx,
 nofootinbib,
]{revtex4-2}

\usepackage{graphicx}
\usepackage{dcolumn}
\usepackage{bm}
\usepackage{physics2}
\usephysicsmodule{ab}
\usephysicsmodule{op.legacy}
\usephysicsmodule{braket}
\usephysicsmodule{nabla.legacy}
\usepackage{derivative,fixdif}
\usepackage{mathtools}
\usepackage[normalem]{ulem}

\usepackage{siunitx}
\usepackage{appendix}
\usepackage{xcolor}
\usepackage[colorlinks,linkcolor=blue,urlcolor=blue, citecolor=blue]{hyperref}
\usepackage[whole]{bxcjkjatype}
\usepackage[T1]{fontenc}
\usepackage{comment}
\usepackage{amsfonts}
\usepackage{svg}
\usepackage{multirow}

\usepackage{newtxtext}
\usepackage{newtxmath}
\usepackage{enumitem}

\definecolor{ssccol}{rgb}{0,0.6,0.8}
\definecolor{nccol}{rgb}{0.6,0.3,0.0}
\definecolor{knkcol}{rgb}{0.3,0.6,0.0}

\begin{document}

\title{Strong coupling of a reconfigurable ${}^{171}$Yb atom array to a tunable telecom-band nanofiber cavity}

\author{H. Ozawa}
\thanks{equal contribution}
\affiliation{Nanofiber Quantum Technologies, Inc. (NanoQT), 1-22-3 Nishiwaseda, Shinjuku-ku, Tokyo 169-0051, Japan}
\author{N. Chen}
\thanks{equal contribution}
\affiliation{Nanofiber Quantum Technologies, Inc. (NanoQT), 1-22-3 Nishiwaseda, Shinjuku-ku, Tokyo 169-0051, Japan}
\author{Y. Hisai}
\affiliation{Nanofiber Quantum Technologies, Inc. (NanoQT), 1-22-3 Nishiwaseda, Shinjuku-ku, Tokyo 169-0051, Japan}
\author{S. Hashimoto}
\affiliation{Nanofiber Quantum Technologies, Inc. (NanoQT), 1-22-3 Nishiwaseda, Shinjuku-ku, Tokyo 169-0051, Japan}
\affiliation{Department of Applied Physics, Waseda University, 3-4-1 Okubo, Shinjuku-ku, Tokyo 169-8555, Japan}
\author{K. N. Komagata}
\affiliation{Nanofiber Quantum Technologies, Inc. (NanoQT), 1-22-3 Nishiwaseda, Shinjuku-ku, Tokyo 169-0051, Japan}
\author{R. T. Oddon}
\affiliation{Nanofiber Quantum Technologies, Inc. (NanoQT), 1-22-3 Nishiwaseda, Shinjuku-ku, Tokyo 169-0051, Japan}
\author{S. Yang}
\affiliation{Nanofiber Quantum Technologies, Inc. (NanoQT), 1-22-3 Nishiwaseda, Shinjuku-ku, Tokyo 169-0051, Japan}
\author{S. Horikawa}
\affiliation{Nanofiber Quantum Technologies, Inc. (NanoQT), 1-22-3 Nishiwaseda, Shinjuku-ku, Tokyo 169-0051, Japan}
\author{S. Kikura}
\affiliation{Nanofiber Quantum Technologies, Inc. (NanoQT), 1-22-3 Nishiwaseda, Shinjuku-ku, Tokyo 169-0051, Japan}
\author{S. Miki}
\affiliation{Advanced ICT Research Institute, National Institute of Information and Communications Technology, 588-2 Iwaoka, Kobe, Hyogo, 651-2492, Japan}
\author{T. Aoki}
\affiliation{Nanofiber Quantum Technologies, Inc. (NanoQT), 1-22-3 Nishiwaseda, Shinjuku-ku, Tokyo 169-0051, Japan}
\affiliation{Department of Applied Physics, Waseda University, 3-4-1 Okubo, Shinjuku-ku, Tokyo 169-8555, Japan}
\affiliation{RIKEN Center for Quantum Computing (RQC), RIKEN, 2-1 Hirosawa, Wako, Saitama, 351-0198, Japan.}
\author{T. Yoshitake}
\affiliation{Nanofiber Quantum Technologies, Inc. (NanoQT), 1-22-3 Nishiwaseda, Shinjuku-ku, Tokyo 169-0051, Japan}
\author{H. Konishi}
\affiliation{Nanofiber Quantum Technologies, Inc. (NanoQT), 1-22-3 Nishiwaseda, Shinjuku-ku, Tokyo 169-0051, Japan}
\author{S. Sunami}
\email{shinichi.sunami@nano-qt.com}
\affiliation{Nanofiber Quantum Technologies, Inc. (NanoQT), 1-22-3 Nishiwaseda, Shinjuku-ku, Tokyo 169-0051, Japan}
\affiliation{Clarendon Laboratory, University of Oxford, Oxford OX1 3PU, United Kingdom}
\author{A. Goban}
\email{akihisa.goban@nano-qt.com}
\affiliation{Nanofiber Quantum Technologies, Inc. (NanoQT), 1-22-3 Nishiwaseda, Shinjuku-ku, Tokyo 169-0051, Japan}
\author{R. Inoue}
\email{ryotaro.inoue@nano-qt.com}
\affiliation{Nanofiber Quantum Technologies, Inc. (NanoQT), 1-22-3 Nishiwaseda, Shinjuku-ku, Tokyo 169-0051, Japan}

\begin{abstract}
Nanophotonic cavities provide strong atom-photon coupling in a compact, directly fiber-coupled geometry, while a reconfigurable array of $^{171}$Yb atoms combines high-fidelity gates with a metastable qubit state that couples directly to a telecom-band optical transition, making the two a natural interface between neutral-atom quantum processors and telecom quantum networks.
Here, we realize strong atom-cavity coupling on the ${}^{3}\mathrm{P}_0 \leftrightarrow {}^{3}\mathrm{D}_1$ transition of ${}^{171}\mathrm{Yb}$, establishing, to our knowledge, the first neutral-atom cavity-QED system to reach the single-atom strong-coupling regime on a telecom-band atomic transition.
Reflection spectroscopy of a single-sided cavity yields a single-atom internal cooperativity of $C_{\mathrm{in}}=5.6(1.0)$ and collective coupling following $\sqrt{N}$ scaling up to $N=5$ atoms, with homogeneous coupling at antinodes over approximately $200\,\upmu\mathrm{m}$ along the nanofiber.
Thermal tuning of a fiber Bragg grating provides in situ control of the cavity outcoupling rate over two orders of magnitude across the undercoupled, critically coupled, and overcoupled regimes, relevant for a wide range of atom-cavity protocols.
We confirm compatibility with the operations of reconfigurable atom arrays by demonstrating high-fidelity imaging with 99.95(6)\% fidelity and trap lifetimes comparable to free-space tweezer array, at $1\,\upmu\mathrm{m}$ from the nanofiber axis. 
Moreover, the diameter-engineered antireflection design of the nanofiber enables tweezer-based atom transport across the nanofiber region with no measurable loss or heating at 4.5\,$\upmu$m above the nanofiber, supporting a two-layer architecture for the nanofiber-integrated operation of the neutral atom array.
From the demonstrated atom-cavity coupling, atom capacity, and cavity mode profile, we project that time-multiplexed remote atom-atom entanglement generation at 40\,kHz would become achievable, with prospects for further scaling through technical enhancements and channel multiplexing.

\end{abstract}
\maketitle

\section{Introduction}
Quantum interconnects play a crucial role in distributing entanglement between
remote matter qubits, enabling a wide range of applications such as secure quantum communication, networked quantum sensing, and distributed quantum computing~\cite{Kimble2008, Wehner2018, Covey2023, menssen2026, Azuma2023}. 
A central requirement for realizing these applications at scale is the generation of remote entanglement at high rates. 
This requirement is particularly stringent for distributed fault-tolerant quantum computing, requiring entanglement generation rates that are orders of magnitude above the demonstrated values at the \(10^2\,\mathrm{s}^{-1}\) level~\cite{Sunami2025perspective,Sinclair2025,sunami2026entboosting}.
At present, these rates are limited by the photon collection efficiency and the time overhead of each entanglement attempt~\cite{OReilly2024, Main2025}. 
Bridging this gap requires both efficient atom-photon interactions and architectures that natively support multiplexed entanglement generation.

\begin{figure*}[t]
    \centering
    \includegraphics[width=\textwidth]{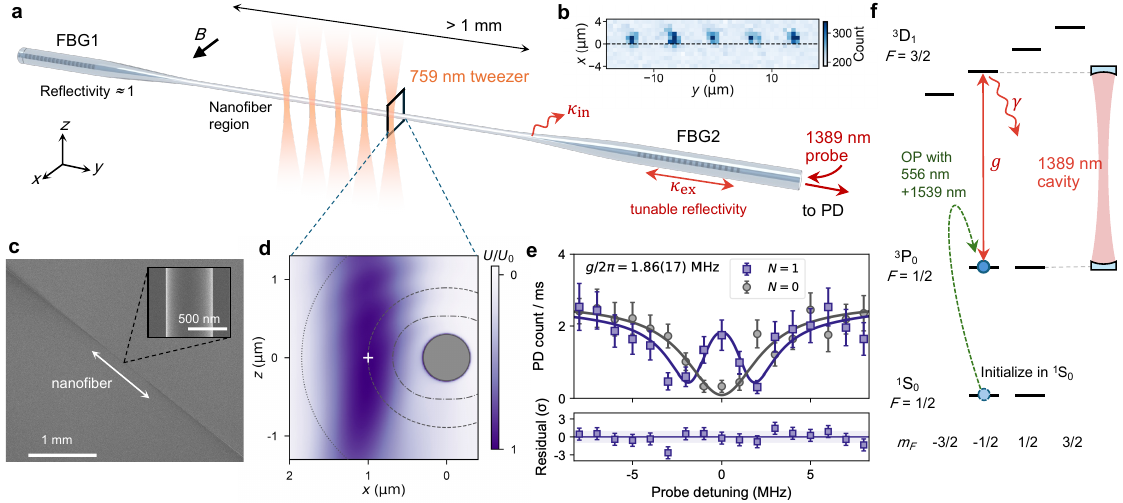}
    \caption{\textbf{Telecom-band nanofiber cavity with an array of $^{171}$Yb atoms in optical tweezers.}
    \textbf{a}, Overview of the platform. The nanofiber cavity is formed by a pair of fiber Bragg gratings (FBG1, FBG2) inscribed in the standard-fiber sections on either side of the nanofiber region, directly connected to the detection system.
    ${}^{171}$Yb atoms are loaded into a 759-nm optical tweezer array and transported to $1\,\upmu$m from nanofiber axis, where the atoms are strongly coupled to the evanescent cavity mode. 
    \textbf{b}, Single-shot raw image of a five-site tweezer array, measured with standard 556-nm fluorescence imaging for ground-state atoms.
    Individual sites remain clearly resolved under illumination by two retroreflected excitation beams propagating in the $y$--$z$ plane, with atoms placed at $1\,\upmu\mathrm{m}$ from the nanofiber axis; the black dashed line denotes the nanofiber axis at $x=0$.
    For illustrative purpose, this image was taken without light-assisted collision, while for all other data presented in this paper we perform it to ensure single occupancies.
    \textbf{c}, Scanning electron micrograph of a fabricated nanofiber, with a nominal diameter of 550\,nm.
    \textbf{d}, Calculated 759-nm tweezer potential in the transverse plane of the nanofiber; the dashed line marks the contour of atom-cavity coupling strength, $g_\mathrm{0}/2\pi = $2.95\,MHz, calculated from the $\mathrm{HE}_{11}$ mode for the 550-nm-diameter nanofiber at the position of the trap center (white cross). 
    Dash-dotted and dotted lines are the contours for $1.7g_\mathrm{0}$ and $0.3g_\mathrm{0}$, respectively.
    \textbf{e}, Cavity-reflection spectra with (purple) and without single atom (gray) loaded to a single-site tweezer next to the nanofiber, where the reported photon-detector (PD) counts are scaled to correct for the frequency-dependent input power (Methods).
    Solid lines are fits to the input-output model, yielding $g/2\pi= 1.86(17)$ MHz. 
    The lower panel shows the residuals of the fit to the single-atom (\(N=1\)) spectrum, with no evident systematic structure.
    \textbf{f}, Relevant ${}^{171}$Yb levels.
    Atoms initialized in \(^{1}\mathrm{S}_{0}\) are optically pumped (OP) to \(^{3}\mathrm{P}_{0}\) using 556- and 1539-nm light and coupled to the cavity whose resonance is tuned to the \(^{3}\mathrm{P}_{0}\leftrightarrow{}^{3}\mathrm{D}_{1}\) transition at $1389\,$nm.
    \label{fig:overview}}
\end{figure*}

Integrating optical cavities with atom arrays can increase the success probability of individual entanglement attempts by enhancing emission into a well-defined optical mode~\cite{Ritter2012} or enabling state-dependent photon scattering~\cite{Duan2004,Reiserer2014}.
The overall rate can then be further increased by multiplexing these attempts in time and in separate channels~\cite{Huie2021,Li2024,Sunami2025perspective, Sinclair2025}.
In a time-multiplexed scheme, multiple atoms are addressed sequentially through a common cavity mode, amortizing atom-transport and initialization overheads over many attempts.
Channel multiplexing instead uses multiple atom-cavity channels in parallel.
Recent experiments with atom arrays in bulk and fiber Fabry-P\'erot cavities have demonstrated programmable and collective atom-cavity coupling, cavity-mediated operations, and site-selective readout toward time-multiplexed operations~\cite{Deist2022mid, Liu2023, Hartung2024, Grinkemeyer2025, Hu2025, Ye2026Chen, Picot2026, Roschinski2026}, while compact nanophotonic resonators and cavity-array microscopes provide routes to scalable channel multiplexing~\cite{Dordevic2021, Menon2024, Zhou2026,Dayan2026, Shaw2026}.
Combining these two forms of multiplexing requires a cavity-based quantum interconnect that unifies these capabilities with a large, individually addressable atom array.

For long-distance quantum networking, a complementary requirement is low-loss photon transmission between remote nodes. 
This favors operation at a telecom band, where propagation loss in silica fiber is low.
Most neutral-atom cavity interfaces operate at wavelengths shorter than the telecom bands, requiring quantum frequency conversion for low-loss fiber transmission~\cite{vanLeent2022, ZhouWeinfurter2024}. 
The \(^{3}\mathrm{P}_{0}\leftrightarrow{}^{3}\mathrm{D}_{1}\) transition of \(^{171}\mathrm{Yb}\) at 1389\,nm provides a direct alternative, connecting the metastable qubit state to telecom-band photons~\cite{Covey2019telecom, Li2024, Sunami2025perspective, Li2025}. 
% Huie2021 uses $^{3}\mathrm{P}_{1}\leftrightarrow{}^{3}\mathrm{D}_{2}$
The \(^{171}\mathrm{Yb}\) platform further combines this transition with the optical-metastable-ground qubit architecture~\cite{Lis2023, Norcia2023}, high-fidelity Rydberg-mediated two-qubit gates~\cite{Muniz2025, Senoo2026, Zhang2026} and efficient implementation of logical qubits via erasure conversion~\cite{Wu2022,Ma2023,Zhang2026}.
High-fidelity atom-photon entanglement with parallelized free-space photon collection has recently been demonstrated on the 1389-nm transition~\cite{Li2025}.
However, strong coupling between neutral atoms and an optical cavity operating directly on a telecom-band transition has not yet been demonstrated.
Reaching this regime would combine cavity-enhanced emission into a single optical mode and cavity-assisted photon-scattering protocols with native telecom-band operation, offering a route to higher-rate and high-fidelity quantum interconnects~\cite{Kikura2025caps,Li2024,Sunami2025perspective}.

Nanofiber cavities are well suited to combine these requirements. 
Their low-loss, uniform-diameter waists can extend over hundreds of micrometers to millimeters, allowing hundreds of atoms to couple to a common evanescent cavity mode. 
Cavity-free nanofiber interfaces have enabled evanescent-field trapping of more than \(10^3\) atoms~\cite{Vetsch2010} and coupling of more than 100 individually addressable, tweezer-trapped atoms to the guided optical mode~\cite{Takahata2026}.
In contrast, strong coupling in nanofiber cavities has so far been demonstrated only with single trapped atoms~\cite{Kato2015,Nayak2019}.
Their subwavelength transverse footprint and direct fiber integration are also compatible with parallel operation across multiple cavities~\cite{Sunami2025perspective}. 
However, integration with a neutral-atom processor requires preserving key atom-array capabilities near the nanofiber, including high-fidelity imaging, long trap lifetimes, and low-loss atom transport across the device.
Together with direct telecom-band operation~\cite{Horikawa2025}, these capabilities provide a route to high-rate quantum interconnects through temporal multiplexing within each cavity and channel multiplexing across multiple cavities.

In this work, we realize strong atom-photon coupling on the telecom-band ${}^{3}\mathrm{P}_0 \leftrightarrow {}^{3}\mathrm{D}_1$ transition of ${}^{171}\mathrm{Yb}$ by integrating a nanofiber cavity with a reconfigurable array of tweezer-trapped atoms, to our knowledge, the first neutral-atom cavity-QED system to reach the strong-coupling regime at telecom wavelengths.
We first establish the cavity-QED functionality of the interface through subwavelength control of the atom-cavity coupling, collective coupling of multiple atoms, and in situ tuning of the cavity outcoupling rate.
We then address integration with neutral-atom quantum processors by demonstrating high-fidelity site-resolved imaging next to the nanofiber and atom transport across the device without measurable loss or heating.
Together, these capabilities establish an array-compatible telecom-band cavity-QED interface for multiplexed neutral-atom quantum interconnects.

\section{Telecom-band nanofiber cavity and Ytterbium atom array}
\label{sec:yb-atoms}

Our platform combines a telecom-band nanofiber cavity with a reconfigurable array of ${}^{171}\mathrm{Yb}$ atoms trapped in 759-nm optical tweezers (Fig.~\ref{fig:overview}a,b,c; see Methods). 
The cavity is formed by two fiber Bragg gratings (FBGs) surrounding a nominally \(550\)-nm-diameter nanofiber waist and is operated in a single-sided cavity configuration at \(\lambda=1389\,\mathrm{nm}\). 
FBG1 serves as the high-reflectivity end mirror with negligible external coupling, whereas FBG2 is the partially reflecting, tunable output coupler that sets the external coupling rate \(\kappa_{\mathrm{ex}}\);
the internal loss rate is \(\kappa_{\mathrm{in}}/2\pi = 1.28(6)\,\mathrm{MHz}\). 
The nanofiber cavity supports an evanescent cavity mode that extends outside the nanofiber surface, while the tweezer array enables individual atoms to be transported within the cavity mode without significantly deforming the trapping potential, with the operating geometry at $1\,\upmu\mathrm{m}$ from the nanofiber axis referred to as the side-trap configuration (Fig.~\ref{fig:overview}d). 
In each experimental cycle, cold atoms are delivered to the science chamber, loaded into the tweezer array with light-assisted collisions, and transported to the side-trap configuration, after which site-resolved fluorescence imaging determines the presence of an atom.
A representative single-shot fluorescence image of a five-site tweezer array shows that the individual sites remain clearly resolved when the array is positioned $1\,\upmu\mathrm{m}$ from the nanofiber axis (Fig.~\ref{fig:overview}b). 
With the chosen excitation geometry, scattering from the nanofiber remains sufficiently suppressed for high-fidelity site-resolved imaging (see also Sec.~\ref{sec:imaging}). 
The atoms are subsequently cooled to approximately $5.5(6)\,\upmu\mathrm{K}$ at the initial loading depth near the nanofiber, at a distance of $1\,\upmu\mathrm{m}$ from the geometrical center of the nanofiber, and optically pumped into the clock state ${}^{3}\mathrm{P}_{0}$, $m_F=-1/2$, at a reduced trap depth (Fig.~\ref{fig:overview}f; see Methods). 

To characterize the atom-cavity coupling, we perform reflection spectroscopy through the FBG2 mirror.
In each experimental realization, a 1-ms weak-probe pulse measures the atom-cavity response at a selected probe detuning, with the reflected photons spectrally filtered and detected using a superconducting-nanowire single-photon detector (see Methods).

Figure~\ref{fig:overview}e shows the atom-cavity reflection spectra with a single optical tweezer placed in the side-trap configuration, with ($N=1$) and without ($N=0$) an atom loaded.
We show the count rate (counts per ms) of the full 1-ms probe for the empty cavity ($N=0$) to provide sufficient statistics, while for the $N=1$ case, we compute the count rate from the first \(250\,\upmu\mathrm{s}\) of the probe pulse since the atomic signal decays with probe duration due to the finite branching from the \({}^{3}\mathrm{D}_{1}\) to the \({}^{3}\mathrm{P}_{1}\) and \({}^{3}\mathrm{P}_{2}\) states.
We fit the data for $N=1$ with a single-sided cavity input-output model, which yields a measured single-atom coupling of $g/2\pi=1.86(17)\,\mathrm{MHz}$ [see Methods, Eq.~\eqref{eq:reflection-coef}].
Using the atomic polarization-decay rate $\gamma/2\pi=0.24\,\mathrm{MHz}$~\cite{Beloy2012} and the internal loss rate $\kappa_{\mathrm{in}}$, the fitted coupling corresponds to a single-atom internal cooperativity of $C_{\mathrm{in}}=g^2/(2\gamma\kappa_{\mathrm{in}})=5.6(1.0)$~\cite{Goto2019}.
The measured coupling is below the value $g_{0}/2\pi=2.95\,\mathrm{MHz}$ ($C_{\mathrm{in,0}}=14.2$) calculated from the $\mathrm{HE}_{11}$ mode profile for an atom at the trap minimum and at an antinode of the intracavity standing wave, a reduction we attribute to the thermal spread of the atom in the tweezer and to imperfect clock-state preparation and depumping during the probe (Methods).

\begin{figure}[t!]
    \centering
    \includegraphics[width=\linewidth]{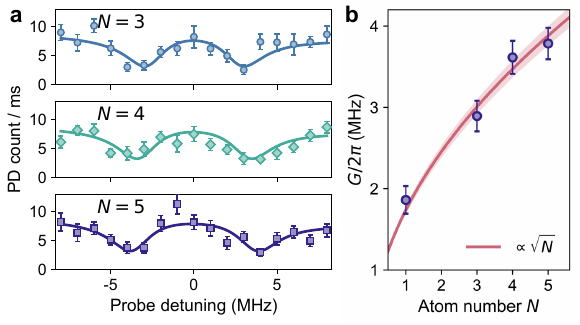}
    \caption{\textbf{Collective atom-cavity coupling with an array of single atoms.} 
    \textbf{a}, Cavity-reflection spectra for $N=3$, 4 and 5 atoms, fitted with the input-output model, yielding collective coupling of $G/2\pi = $ 2.89(19), 3.61(20), and 3.78(19)\,MHz, respectively. 
    We used higher probe power than the data shown in Fig.~\ref{fig:overview}e.
    \textbf{b}, The fitted values of $G/2\pi$ as a function of atom number $N$, together with the measured single-atom value from Fig.~\ref{fig:overview}e, as well as a fitted scaling $G/2\pi = G_1\sqrt{N}$ where the shading represents the standard error of the fitted prefactor $G_1 = 1.74(5)$\,MHz.
    \label{fig:collective}}
\end{figure}

\begin{figure*}[t]
    \centering
    \includegraphics[width=0.6\linewidth]{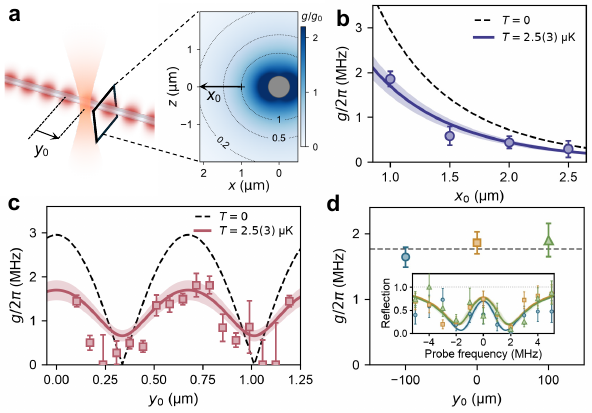}
    \caption{\textbf{Spatially resolved atom-cavity coupling.}
    \textbf{a}, Geometry of the position scans relative to the nanofiber cavity mode.
    The coordinates $x_0$ and $y_0$ denote radial displacement and translation along the nanofiber axis, respectively.
    The inset shows the calculated atom-cavity coupling strength normalized to its value at the operating position $(x,z)=(1,0)\,\upmu\mathrm{m}$ at the antinode of the standing-wave pattern along $y$.
    \textbf{b}, The atom-cavity coupling rate obtained from the cavity reflection spectra measured at varying $x_0$, resolving the exponential decay of the evanescent cavity field.
    The dashed line is the ideal model calculated from the $\mathrm{HE}_{11}$ mode profiles at zero temperature and with perfect state preparation, while the solid curve is a model at temperature $T=2.5(3)\,\upmu$K, with the shaded region representing uncertainty of the model (Methods).
    \textbf{c}, Measured coupling as a function of axial position $y_0$, resolving the intracavity standing-wave structure at the subwavelength scale.
    The model is the same as in \textbf{b}, with only the origin of $y_0$ adjusted to the antinode.
    \textbf{d}, Atom-cavity coupling measured at antinodes near $y_0=-100$, $0$, and $100\,\upmu\mathrm{m}$, showing comparable peak coupling over $200\,\upmu\mathrm{m}$.
    Inset shows representative cavity-reflection spectra at the three axial positions normalized by their far-detuned counts, with fitted spectrum shown for each data in solid curves.}
    \label{fig:coupling}
\end{figure*}

Next, we investigate the collective response of single atoms trapped in multiple tweezer sites coupled to the shared cavity mode. 
To realize comparable coupling magnitudes across the array, we set the tweezer spacing to around \(8.8\,\upmu\mathrm{m}\), close to an integer multiple of the intracavity antinode spacing $\lambda/(2n_{\mathrm{eff}})$, where $n_{\mathrm{eff}}=1.027$ is the effective refractive index for the fiber diameter of 550\,nm.
Figure~\ref{fig:collective}a shows cavity-reflection spectra for $N=3$, 4 and 5 atoms, with the normal-mode splitting increasing with atom number. 
Fits to each spectrum for each atom number $N$ yield the collective coupling $G=(\sum_i g_i^2)^{1/2}$. 
The obtained values of $G$, together with the single-atom coupling $g$ from Fig.~\ref{fig:overview}e, follow the expected $G=\sqrt{N}g$ scaling (Fig.~\ref{fig:collective}b), demonstrating the compatibility of our cavity with an atom array.

\section{Sub-wavelength probe of evanescent cavity mode via atom-cavity spectroscopy}\label{sec:coupling}

Optical tweezers enable direct control of the atom-cavity coupling by positioning individual atoms within the evanescent cavity mode.
To characterize its spatial dependence, we use a single tweezer containing a single atom and scan the radial position $x_0$ and the position $y_0$ along the nanofiber axis, with the remaining coordinates fixed at their operating values (Fig.~\ref{fig:coupling}a).
The measured coupling along $x_0$ exhibits the characteristic evanescent-field decay of the cavity field (Fig.~\ref{fig:coupling}b), while the scan along $y_0$ resolves the intracavity standing-wave structure at the subwavelength scale (Fig.~\ref{fig:coupling}c).
The solid curves show the coupling expected for an atom at finite temperature, including the state-preparation and depumping corrections, with the origin of $y_0$ fitted from this data (Methods); the reduced standing-wave contrast relative to the ideal $T=0$ model (dashed) reflects the thermal spread of the atom along the nanofiber axis.
The observed period agrees with the expected intracavity standing-wave period $\lambda/(2n_{\mathrm{eff}})=676\,\mathrm{nm}$.
To assess the spatial uniformity of the coupling across the nanofiber region, we further measure the atom-cavity coupling at antinodes near $y_0=-100$, $0$, and $100\,\upmu\mathrm{m}$.
The extracted peak couplings are comparable across these positions (Fig.~\ref{fig:coupling}d), demonstrating access to similar antinode coupling over approximately $200\,\upmu\mathrm{m}$.
The reachable $y_0$ is currently limited by the finite field of view of the objective lens;
scanning electron micrograph measurements of the nanofiber diameter across the nanofiber region of 1\,mm for the separately fabricated device show less than 1\% fluctuation of the diameter, suggesting that the uniformity of the atom-photon coupling extends to a millimeter scale, since the $\mathrm{HE}_{11}$ mode profiles are determined by the fiber diameter and wavelength (Fig.~\ref{fig:SEM}).
For a realistic spatial separation of tweezer sites of 4\,$\upmu$m, and with both sides of the nanofiber usable in the side-trap configuration, the demonstrated uniformity over 200\,$\upmu$m shows the capacity of 100 tweezer sites with homogeneous atom-cavity coupling, which extends to 500 sites for a 1\,mm uniformity region.

\section{Tunability of external coupling rate}\label{sec:tuning}

In conventional optical cavities, the outcoupling rate $\kappa_{\mathrm{ex}}$ is a parameter fixed at the time of cavity fabrication, limiting their application to a fixed set of protocols. 
In situ control of $\kappa_{\mathrm{ex}}$ not only broadens the applicability of an atom-cavity system but also enables fine tuning of parameters required to achieve high-fidelity operations;
for example, cavity-mediated spin-spin interactions and many-body entangled-state preparation improve monotonically as $\kappa_{\mathrm{ex}} \to 0$~\cite{Li2022spinspin, Ramette2022anytoany}, while critical coupling $\kappa_{\mathrm{ex}} = \kappa_{\mathrm{in}}$ results in high contrast in reflectivity and enables single-atom optical switching and routing~\cite{Aoki2006, OShea2013, Will2021}.
In the overcoupled regime, $\kappa_{\mathrm{ex}}>\kappa_{\mathrm{in}}$, high-fidelity cavity-assisted photon-scattering gates require $\kappa_{\mathrm{ex}}=\sqrt{1+2C_{\mathrm{in}}}\,\kappa_{\mathrm{in}}$, where the atom-coupled and empty-cavity reflection amplitudes are matched~\cite{Duan2004,Goto2010,Kikura2025caps,Grinkemeyer2025}.
At larger $\kappa_{\mathrm{ex}}$, stronger overcoupling maximizes photon extraction and suppresses the photon impurity that limits emission-based protocols~\cite{Li2024,Kikura2025recoil}.

\begin{figure}[t!]
    \centering
    \includegraphics[width=\linewidth]{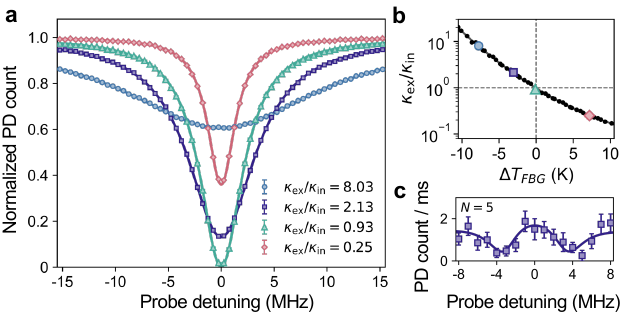}
    \caption{
    \textbf{In situ tuning of the external coupling rate.}
    \textbf{a}, Measured empty cavity reflection spectra at $\kappa_{\mathrm{ex}}/\kappa_{\mathrm{in}}=8.03(42)$, $2.13(15)$, $0.93(9)$, and $0.25(6)$, covering the overcoupled, near-critically coupled, and undercoupled regimes.
    \textbf{b}, Control of the external coupling rate $\kappa_{\rm ex}/\kappa_{\rm in}$ by tuning the FBG2 temperature offset $\Delta T_{\rm FBG}$.
    The colored markers indicate the operating points corresponding to the spectra in \textbf{a}; the dashed lines mark critical coupling.
    \textbf{c}, Representative cavity-reflection spectrum with $N=5$ atoms at $\kappa_{\rm ex}/\kappa_{\rm in}=1.82(9)$ with fitted collective atom-cavity coupling of $G/2\pi=3.9(1.2)$\,MHz, demonstrating that atom-cavity coupling is maintained in the overcoupling regime.
    \label{fig:tunability}}
\end{figure}

Nanofiber cavities formed by thermally tunable fiber Bragg grating mirrors, analogous to whispering-gallery-mode resonators with tunable taper coupling~\cite{Aoki2006,OShea2013}, allow \(\kappa_\mathrm{ex}\) to be tuned in situ~\cite{Kato2015}.
We demonstrate this control by shifting the stopband of the outcoupling FBG2 mirror while maintaining the cavity resonance and internal loss rate approximately fixed.
As shown in Fig.~\ref{fig:tunability}b, varying the FBG2 temperature tunes $\kappa_\mathrm{ex}/\kappa_\mathrm{in}$ over nearly two orders of magnitude, spanning the undercoupled, critically coupled, and overcoupled regimes.
The FBG2 temperature is inferred from the measured stopband shift using the thermo-optic coefficient of silica and the thermal expansion of the grating period.
Representative empty-cavity reflection spectra at four coupling conditions are shown in Fig.~\ref{fig:tunability}a.
At the undercoupled end of this tuning range, setting $\kappa_{\mathrm{ex}}\simeq0.1\kappa_{\mathrm{in}}$ reduces the total cavity decay rate to $\kappa_{\mathrm{tot}}/2\pi=(\kappa_{\mathrm{in}}+\kappa_{\mathrm{ex}})/2\pi\simeq1.41\,\mathrm{MHz}$, allowing access to the single-atom strong-coupling condition $g>\kappa_{\mathrm{tot}},\gamma$ for the measured coupling strength~\cite{Reiserer2015}.
In the overcoupled regime, a cavity-reflection spectrum with $N=5$ atoms is shown in Fig.~\ref{fig:tunability}c for $\kappa_\mathrm{ex}/\kappa_\mathrm{in}=1.82(9)$, demonstrating that the atom-cavity coupling can be maintained for a range of cavity parameters.
The accessible tuning range includes the photon-scattering optimum,
$\kappa_\mathrm{ex}=\sqrt{1+2C_\mathrm{in}}\,\kappa_\mathrm{in}\approx3.5\,\kappa_\mathrm{in}$
for our measured $C_\mathrm{in}$, and extends further into the overcoupled regime relevant to efficient photon emission, allowing the same cavity to access distinct protocol-specific operating conditions without reinstallation or modification of the device.

\section{Reconfigurable atom array next to the nanofiber} \label{sec:imaging}

\begin{figure*}[t]
    \centering
    \includegraphics[width=0.75\linewidth]{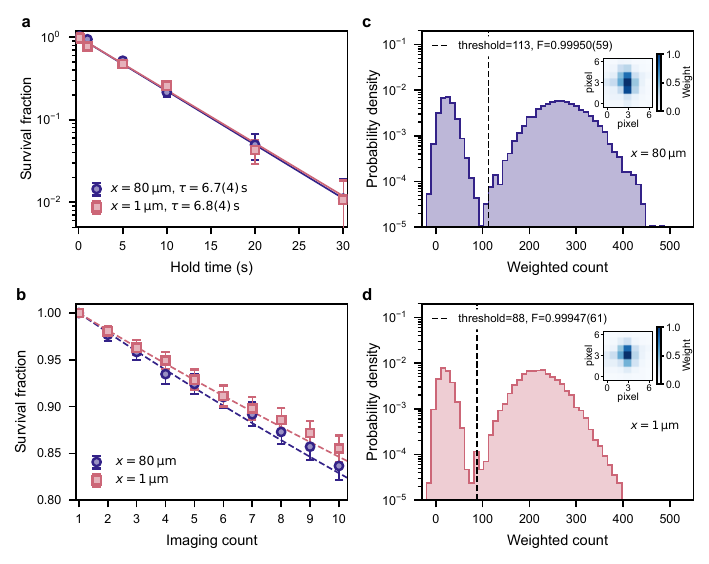}
    \caption{\textbf{Operating tweezer-trapped atom array next to a nanophotonic device.}
    \textbf{a}, Lifetime of the trapped atoms. 
    Survival fraction of tweezer-trapped atoms as a function of hold time in free space ($x = 80\,\upmu$m) and at the side-trap configuration ($x = 1\,\upmu$m).
    Exponential fits yield the lifetimes of \(\tau=6.7(4)\,\mathrm{s}\) and \(6.8(4)\,\mathrm{s}\), respectively.
    \textbf{b}, 
    Fraction of sites bright in the first image that remain bright in imaging count $j$, with error bars denoting bootstrap 95\% intervals. 
    Dashed lines show $p_\text{s}^{j-1}$ with the per-image survival probability $p_{\mathrm{s}}=97.9(4)\%$ ($x=80\,\upmu\mathrm{m}$) and $98.2(4)\%$ ($x=1\,\upmu\mathrm{m}$) obtained from the three-image analysis (see Methods).
    \textbf{c},\textbf{d}, Histograms of the weighted counts of the three-image analysis at $x = 80\,\upmu$m (\textbf{c}) and $x = 1\,\upmu$m (\textbf{d}); insets show the empirical point-spread-function weight maps used to obtain weighted counts (Methods).
    Dashed lines mark the thresholds that maximize the imaging fidelity, $F_\mathrm{img} = 99.95(6)\,\%$ at both positions. 
    \label{fig:imaging}}
\end{figure*}

Operating tweezer-trapped atom arrays in the vicinity of the nanofiber requires that the atoms remain stably trapped there.
We first verify this by comparing the atom lifetime near and far from the nanofiber.
For the side-trap configuration, the atom lifetime is $6.8(4)\,\mathrm{s}$, consistent with the value of $6.7(4)\,\mathrm{s}$ measured at $x=80\,\upmu\mathrm{m}$, demonstrating stable trapping without a measurable reduction in lifetime near the nanofiber (Fig.~\ref{fig:imaging}a).

Beyond stable trapping, fluorescence imaging of tweezer-trapped atoms enables the parallel, high-fidelity readout required to operate large-scale atom arrays. 
However, near photonic structures, light from the excitation beams scatters off the structure and produces a background that obscures the atomic fluorescence. 
Here, we suppress this background while retaining fluorescence imaging directly on the 556-nm transition by using only two of the three retroreflected excitation-beam axes. 
Specifically, the excitation beams propagating in the $y$--$z$ plane are kept incident on the atoms, while the beam along the orthogonal $x$ axis is blocked, with the coordinate axes defined in Fig.~\ref{fig:overview}a. 
This geometry strongly suppresses nanofiber-scattered light entering the imaging system while retaining sufficient fluorescence signal for site-resolved detection (see Methods). 

We next quantitatively characterize the imaging performance using single atoms in the array.
With a 30-ms imaging pulse, repeated imaging yields per-image survival probabilities of $97.9(4)\%$ at $x=80\,\upmu\mathrm{m}$ and $98.2(4)\%$ at $x=1\,\upmu\mathrm{m}$ (side-trap configuration; Fig.~\ref{fig:imaging}b). 
The corresponding single-atom fluorescence distributions remain well separated at both positions, yielding an imaging fidelity of $99.95(6)\%$ (Fig.~\ref{fig:imaging}c,d; see Methods).
Together, these results demonstrate that atoms can be held and read out directly near the nanofiber cavity without a significant loss of imaging performance relative to the far-from-fiber configuration.

Background-free multilevel fluorescence imaging provides an alternative approach by spectrally separating the detected fluorescence from the excitation light, as demonstrated for Cs atoms near a nanophotonic waveguide~\cite{Menon2024}, and such a scheme is also applicable to Yb. 
In contrast, the present approach suppresses the scattering background through the excitation geometry. 
This provides flexibility in the choice of imaging transition and could, for example, allow the same strategy to be applied to the broad 399-nm ${}^{1}\mathrm{S}_{0}$--${}^{1}\mathrm{P}_{1}$ transition, offering a route toward substantially faster fluorescence imaging near the nanofiber~\cite{Falconi2025}.

\section{Lossless atom transport across the nanofiber}\label{sec:transport}

\begin{figure*}[ht]
    \centering
    \includegraphics[width=0.75\textwidth]{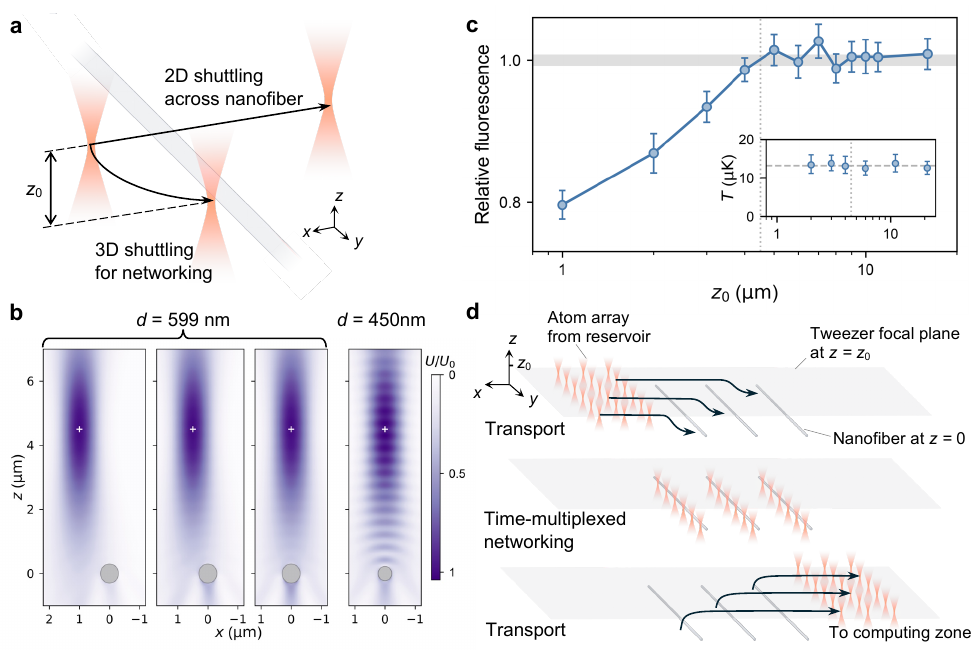}
    \caption{\textbf{Lossless transport across the nanofiber.}
    \textbf{a}, Conceptual transport scheme near the nanofiber, using three-dimensional shuttling to access the cavity mode and a displaced two-dimensional transport layer for transport across the nanofiber with negligible perturbation. 
    \textbf{b}, Calculated tweezer potentials near the nanofiber for representative fiber diameters. At the optimized diameter of \(d=599\,\mathrm{nm}\), the modulation to the tweezer potential is negligible with the tweezer focus \(z_0 = 4.5\,\upmu\mathrm{m}\) from the fiber center, in contrast to the pronounced standing-wave modulation at other diameters such as \(d=450\) nm shown on the right.
    \textbf{c}, Measured atom fluorescence after transport across the nanofiber as a function of \(z_0\). 
    The gray shading at relative fluorescence of 1 corresponds to the fluorescence counts recorded after the same shuttling trajectory in the absence of the effect of the nanofiber (the height of the shading corresponds to the standard deviation of the recorded fluorescence), demonstrating lossless transport above $z_0=4.5\, \upmu\mathrm{m}$.
    The inset further confirms negligible transport-induced heating, measured by release-and-recapture thermometry performed after the transport.
    \textbf{d}, Schematic of a time-multiplexed quantum-networking operation combining parallel transport of atom arrays to multiple nanofiber cavities, sequential entanglement generation attempts for atom arrays coupled to cavities~\cite{Sunami2025perspective,Li2024,Huie2021}, and subsequent transport away from the cavities to the computing zone; the parallel operation of a large atom array amortizes the time cost required for atom shuttling, resulting in high-rate remote entanglement generation across fiber channels; see text.
    \label{fig:transport}}
\end{figure*}

A distinct feature of neutral atom arrays is the parallel atom rearrangement that enables efficient all-to-all connectivity~\cite{Bluvstein2022}.
Retaining this capability alongside a nanophotonic interface, so that atom arrays can be shuttled freely across and between cavities, is what makes multiplexed quantum networking possible and enables integration with neutral-atom quantum information processors.
The requirement is that the trapping light scattered at the dielectric surface, which would otherwise imprint a standing-wave modulation on the tweezer potential, be kept well below the level that disturbs the atomic motion.

The subwavelength, extremely smooth profile of the nanofiber devices, as well as stable and precise control over their diameter (see Fig.~\ref{fig:SEM}), allows our device to satisfy a condition for strong suppression of tweezer-light backscattering via destructive interference of multiple scattering paths (see Methods).
This enables a 3D-atom-shuttling architecture where the standard 2D atom array is supplemented by another layer reserved for quantum networking (Fig.~\ref{fig:transport}a), with layer separation that can be as small as a few micrometers, as we show below.
For such a small separation, recently demonstrated three-dimensional atom shuttling using a pair of acousto-optic deflectors suffices to transfer atoms between the 2D plane and the side-trap configuration with small time overhead~\cite{picard2025,guo2025_3d,Lu26astigmatism}.

To illustrate the suppressed backward reflection, in Fig.~\ref{fig:transport}b, we show the optical tweezer potential calculated from the finite-difference time-domain simulation of the focused Gaussian beam in the vicinity of the infinite dielectric cylinder (nanofiber, gray circle). 
At an optimal diameter of \(d=599\,\mathrm{nm}\) for the 759-nm optical tweezer, the optical potential shows negligible perturbation even when the tweezer focus $z_0$ is only 4.5\,$\upmu$m above the nanofiber, while different fiber diameters result in a substantial standing-wave pattern in an otherwise identical configuration.
This shows that the tweezer potential remains invariant as it passes across the nanofiber device at small $z_0$.
Such strong suppression of the tweezer-light back reflection is achieved only by selecting the appropriate ratio between the nanofiber diameter $d$ and tweezer wavelength $\lambda_\mathrm{tweezer}$, such as $d/\lambda_\mathrm{tweezer} \approx 0.45$ and 0.79 (see Fig.~\ref{fig:reflection-model-comparison}). Therefore, the antireflection design is applicable to a wide range of atomic species and tweezer-light wavelengths (see Fig.~\ref{fig:diameter-design}).

To demonstrate such an architecture, we perform atom transport across the nanofiber using a second, nearly identical experimental system with a nanofiber cavity targeted to be 599\,nm in diameter.
We characterize atom survival after transporting a tweezer across the nanofiber at the speed of 0.14\,$\upmu$m/$\upmu$s, with varying transport-layer displacement \(z_0\). 
We record the atom fluorescence after shuttling and compare the resulting fluorescence signal, as shown in Fig.~\ref{fig:transport}c.
The atom survival probability is consistent with unity for \(z_0\gtrsim4.5\,\upmu\mathrm{m}\), while the atomic temperature remains unchanged, indicating no measurable transport-induced loss or heating.
This enables a parallel transport of atom arrays not just across but to multiple nanofiber cavities, enabling channel- and time-multiplexed entanglement generation attempts (Fig.~\ref{fig:transport}d), with an extended protocol resulting in as small as 40\,$\upmu$s time overhead to bring a new set of atoms to the cavity mode (Fig.~\ref{fig:timemux}).
Such a highly multiplexed operation is a prerequisite for high-rate quantum networking~\cite{Sunami2025perspective, Li2024, Huie2021, Sinclair2025}.
For example, with internal cooperativity $C_\mathrm{in}=5.6$, an atom capacity of 50 per one side of the nanofiber, as well as an atom shuttling time overhead of 40\,$\upmu$s, we estimate the entanglement generation rate of 40\,kHz with the time-multiplexed cavity-assisted photon scattering protocol in Ref.~\cite{Kikura2025caps} at the target Bell pair fidelity of 99\% (Methods).

\section{Conclusion}

We have realized a cavity interface between tweezer-trapped ${}^{171}$Yb atoms and a nanofiber cavity operating directly on the telecom-band ${}^{3}\mathrm{P}_0 \leftrightarrow {}^3\mathrm{D}_1$ transition, achieving a single-atom internal cooperativity of $C_\mathrm{in} = 5.6(1.0)$  and collective coupling with the characteristic $\sqrt{N}$ scaling up to $N=5$ atoms.
To our knowledge, this establishes the first neutral-atom cavity-QED system to reach the single-atom strong-coupling regime on a telecom-band atomic transition.
Using the atoms as local probes, we resolve the evanescent and standing-wave structure of the cavity mode at the subwavelength scale and confirm homogeneous antinode coupling over approximately 200\,$\upmu$m of the nanofiber, providing  capacity for more than 100 tweezer-trapped atoms.
The FBG mirrors provide in situ control of the outcoupling rate over nearly two orders of magnitude, spanning regimes relevant to photon-scattering and photon-emission protocols.
At the same time, the atom array retains its full functionality beside the device: site-resolved imaging with $99.95(6)\%$ fidelity at 1\,$\upmu$m from the nanofiber axis, trap lifetimes comparable to free space, and transport across the nanofiber without measurable loss or heating through diameter-engineered suppression of tweezer-light reflection.
Together, these capabilities provide a basis for combining temporal multiplexing within each cavity and channel multiplexing across multiple cavities in a telecom-band quantum interconnect compatible with large, individually addressable atom arrays.

Several near-term improvements will further strengthen the interface.
Improved atom cooling~\cite{Jenkins2022} and shorter probe duration will resolve most of the observed deviation of the atom-photon coupling from the predicted value $g_0/2\pi = 2.95$\,MHz, reaching $C_\mathrm{in,0}=14.2$.
A shorter tweezer wavelength, such as 486\,nm~\cite{Ma2022}, allows the atom-surface distance to be 450\,nm and further increases the achievable \(C_\mathrm{in}\) to 40. 
Reducing internal loss through improved fabrication, as recently demonstrated in Ref.~\cite{Horikawa2025}, would then raise \(C_\mathrm{in}\) to 65.
The sub-percent diameter variation
over the millimeter-scale nanofiber region enables the extension of the uniformly coupled atom array beyond the present field-of-view-limited \(200\,\upmu\mathrm{m}\) region.

With the seamless integration of neutral-atom quantum computing with ${}^{171}$Yb atoms~\cite{Ma2023,Zhang2026,Senoo2026,atom2026} through lossless atom transport, the demonstrated and near-term performance of the telecom-band nanofiber cavity will form the basis for high-rate, high-fidelity remote entanglement generation~\cite{Sunami2025perspective}, opening up a path towards large-scale distributed fault-tolerant quantum computing, long-distance quantum networking, and beyond.

\section*{Acknowledgements}
We thank V. Vuleti\'c for helpful discussions and the members of the NanoQT team for their assistance.
In particular, we thank S. Kato for his contributions during the early stages of this work, K. Harada for his technical contributions, and E. Otsuka and E. Ohmori for scanning electron micrograph measurements.
We also thank the DeltaFiber team for their contributions to nanofiber-cavity fabrication.

This work was supported by JST Moonshot R\&D (Grant Number JPMJMS2268).
This paper is based on results obtained from a project, JPNP20017, subsidized by the New Energy and Industrial Technology Development Organization (NEDO).

\section*{Author contributions}
The NanoQT team designed, built, and operated the experiment and analyzed the data. 
S.S. conceived the lossless transport scheme, which is the subject of U.S. patent application No. 19/793,903. 
S.M. designed and provided the superconducting nanowire single-photon detector. 
All authors discussed the results and contributed to writing the manuscript.

\section*{Competing interests}
T.A. and A.G are co-founders and shareholders of Nanofiber Quantum Technologies, Inc. 
All the other authors affiliated with Nanofiber Quantum Technologies, Inc are employees.

\bibliography{refs.bib}

\clearpage
\appendix
\section*{Methods}

\subsection{Nanofiber cavity fabrication and frequency stabilization}
\label{md:fabandlock}

The nanofiber cavity is fabricated using the procedure described in Ref.~\cite{Horikawa2025}, namely by tapering a standard silica optical fiber to a nominal waist diameter of $550\,$nm between two FBGs, thereby forming a Fabry-P\'erot cavity resonant near $1389\,$nm.
Scanning electron microscopy of representative nanofibers fabricated with the pulling parameters used for the present device shows a uniform-diameter region extending over $200\,\upmu\mathrm{m}$, while a modified fabrication design extends the uniform-diameter region beyond 1\,mm (See Methods~\ref{md:uniformity}).
The two FBGs serve as a high-reflectivity mirror ($R_1\sim99.98\%$)  and a partially transmitting output coupler ($R_2\sim99.5\%$), defining a cavity with a physical length of 3\,cm and a measured free spectral range (FSR) of $3.17$\,GHz.
Fits to cavity spectra without atoms across the FBG tuning range yield an internal loss rate of $\kappa_{\mathrm{in}}/2\pi=1.28(6)\,\mathrm{MHz}$ and a corresponding intrinsic finesse of $1240(60)$, while the external coupling rates through the high-reflectivity and output mirrors are $\kappa_{\mathrm{ex},1}/2\pi\lesssim 0.05\,\mathrm{MHz}$ and $\kappa_{\mathrm{ex},2}/2\pi=0.1\text{--}10\,\mathrm{MHz}$, respectively.
Because \(\kappa_{\mathrm{ex},1}\ll\kappa_{\mathrm{ex},2}\), throughout the main text we neglect the residual coupling through FBG1 and use \(\kappa_{\mathrm{ex},1}=0\) and \(\kappa_{\mathrm{ex},2}\equiv\kappa_{\mathrm{ex}}\) for simplicity.
Based on the measured FSR and calculated transverse mode profile, the projected single-atom coupling on the ${}^{3}\mathrm{P}_{0},F=1/2,m_F=-1/2\leftrightarrow{}^{3}\mathrm{D}_{1},F'=3/2,m_{F'}=-1/2$ transition is $g_{0}/2\pi=2.95\,\mathrm{MHz}$ for an atom positioned at an antinode of the cavity mode, $1\,\upmu\mathrm{m}$ from the nanofiber axis.

The external coupling rate $\kappa_{\mathrm{ex},2}$ and the cavity resonance frequency are controlled primarily by the temperatures of FBG2 and FBG1, respectively, while guided-mode photothermal heating provides the fine resonance control required for active stabilization.
At the atomic transition frequency, FBG2 operates on the slope of its stopband~\cite{Kato2015}, making $\kappa_{\mathrm{ex},2}$ sensitive to its temperature, whereas FBG1 remains within the high-reflectivity region of its stopband, allowing temperature tuning of its reflection phase to compensate shifts of the cavity resonance with negligible change in $\kappa_{\mathrm{ex},1}$.
Fine resonance tuning is achieved by varying the power of 759-nm light guided through the nanofiber, which changes the cavity optical path length through the nanofiber's photothermal response~\cite{Wuttke2013}.
For active stabilization, an auxiliary locking laser is coupled to a longitudinal cavity mode separated from the target mode by one FSR and provides an error signal that is fed back to the 759-nm heating power, maintaining the target cavity mode on resonance with the atomic transition within 0.1\,MHz. 

\subsection{Atom array preparation}
\label{md:atomprep}
Cold ${}^{171}\mathrm{Yb}$ atoms are delivered to the science chamber housing the nanofiber cavity using the pulsed atom-delivery system described in Ref.~\cite{Hashimoto2026}.
Atoms collected in a 399-nm two-dimensional magneto-optical trap (MOT) approximately $240\,\mathrm{mm}$ from the center of the science chamber are launched using 399-nm moving molasses.
After a ballistic time-of-flight, the $399\text{-}\mathrm{nm}$ beams are reapplied near the science chamber to decelerate the atomic pulse, and the 556-nm three-dimensional MOT is turned on to capture the transported atoms.
The launch and deceleration parameters are optimized to maintain efficient capture while minimizing transverse heating and spatial expansion due to photon scattering on the broad 399-nm transition.
The captured atoms are then compressed and loaded into a 759-nm optical-tweezer array at a tweezer depth of $U/k_B\simeq500\,\upmu\mathrm{K}$, where light-assisted collisions remove the excess atoms from the tweezers. The final tweezer occupancy is then measured by fluorescence imaging on the 556-nm ${}^{1}\mathrm{S}_{0},F=1/2\leftrightarrow{}^{3}\mathrm{P}_{1},F=3/2$ transition with the bias magnetic field set near the magic angle relative to the tweezer polarization to suppress differential light shifts (Fig.~\ref{fig:geometry}b), yielding a single-atom occupation probability of $75\%$ (see Fig.~\ref{fig:imaging}c,d)~\cite{Lis2023}. 

Before proceeding with subsequent experimental steps, the atoms are first transported to the side-trap configuration. 
We use two independent experimental systems with different tweezer-generation and transport capabilities.
In the first system, a spatial light modulator generates the tweezer array below the nanofiber, and transport is performed by translating the objective lens to shift the tweezer focal plane to the nanofiber axis.
In the second system, two orthogonally mounted acousto-optic deflectors (AODs) generate the tweezer array, and transport is performed by chirping the RF tones sent to the AODs.
The first system is used for the atom-cavity measurements and imaging in Figs.~\ref{fig:overview}–\ref{fig:imaging}, whereas the second is used for the transport measurements in Fig.~\ref{fig:transport}.
The individual tweezers in both systems have beam waists of $1\,\upmu\mathrm{m}$, allowing atoms to be positioned at $x=1\,\upmu\mathrm{m}$ without significant deformation of the trapping potential (see Fig.~\ref{fig:overview}d).
Owing to the minimal scattering of the tweezer light by the nanofiber in the side-trap configuration, the measured atom lifetime near the nanofiber is  $6.8(4)\,\mathrm{s}$, consistent with that measured in free space and likely limited by technical factors such as background-gas collisions and tweezer-induced heating. 

After transport to the side-trap configuration, the atoms undergo an additional stage of gray-molasses cooling at the same trap depth used for loading, reaching a temperature of $5.5(6)\,\upmu\mathrm{K}$; the trap depth is then lowered adiabatically over 20\,ms to $U/k_B\simeq100\,\upmu\mathrm{K}$, resulting in a temperature of $T=2.5(3)\,\upmu\mathrm{K}$, which corresponds to mean vibrational quantum numbers of $\bar n\approx2$ and $13$ in the transverse (21\,kHz) and axial (3.7\,kHz) directions of the tweezer.
Next, the atoms are polarized into the ${}^{1}\mathrm{S}_{0}$, $m_F=-1/2$ nuclear spin state by optical pumping via the ${}^{3}\mathrm{P}_{1}$, $F=1/2$ manifold with $99.6(2)\%$ efficiency. 
They are subsequently optically pumped into ${}^{3}\mathrm{P}_0,\,m_F=-1/2$ using simultaneous excitation on the 556-nm ${}^{1}\mathrm{S}_0\rightarrow{}^{3}\mathrm{P}_1,\,F=3/2$ and 1539-nm ${}^{3}\mathrm{P}_1,\,F=3/2\rightarrow{}^{3}\mathrm{D}_1,\,F=3/2$ transitions.
The initial spin-polarizing 556-nm light is retained during this process to repolarize atoms returning to the unwanted ground-state spin through ${}^{3}\mathrm{D}_1\rightarrow{}^{3}\mathrm{P}_1\rightarrow{}^{1}\mathrm{S}_0$ decay (Fig.~\ref{fig:geometry}a,c).
A bias magnetic field of $B = 20\,\mathrm{G}$ defines the quantization axis.
Population excited to the ${}^{3}\mathrm{D}_{1}$ state decays into the ${}^{3}\mathrm{P}_{J}$ manifold. 
Decay into ${}^{3}\mathrm{P}_{0}$ preferentially accumulates population in the target $m_F=-1/2$ state, whereas population decaying into ${}^{3}\mathrm{P}_{1}$ subsequently returns to the ${}^{1}\mathrm{S}_{0}$ ground-state manifold, where it is repolarized into the ${}^{1}\mathrm{S}_{0}$, $m_F=-1/2$ state and re-enters the optical-pumping cycle.
Population decaying into ${}^{3}\mathrm{P}_{2}$ is rapidly lost from the tweezer because this state experiences a repulsive potential~\cite{Hohn2023}.
The clock-state preparation efficiency is constrained by two measurements. 
In the first measurement, we measure the population remaining in the bright ${}^{1}\mathrm{S}_{0}$ state after optical pumping into the ${}^{3}\mathrm{P}_{0}$ state, yielding a residual population of $6(1)\%$, placing an upper bound of $94(1)\%$ on the \({}^{3}\mathrm{P}_{0}\) population because the depleted fraction can also include atoms lost through the anti-trapped \({}^{3}\mathrm{P}_{2}\) state.
In the second measurement, an additional 1389-nm repump beam applied after the optical pumping stage returns \(87(1)\%\) of the initial population from \({}^{3}\mathrm{P}_{0}\) to \({}^{1}\mathrm{S}_{0}\), providing a lower bound because leakage into \({}^{3}\mathrm{P}_{2}\) during repumping reduces the recovered fraction.

\subsection{Fluorescence imaging near the nanofiber and analysis}
\label{md:imaging}
Atoms are imaged in place near the nanofiber on the 556-nm ${}^{1}\mathrm{S}_{0}\leftrightarrow{}^{3}\mathrm{P}_{1},F=3/2$ transition with a 30-ms exposure. 
Two retroreflected excitation beams propagate in the plane defined by the nanofiber and tweezer axes (Fig.~\ref{fig:geometry}a,b).

Images are corrected for the camera dark level, a scalar offset measured with the sensor blocked at the same exposure and readout settings~\cite{Manetsch2025}.
The signal of each site is the sum of camera counts in a $7\times7$-pixel box weighted by a positive map $W$ shared by all sites and images following the procedure in Refs.~\cite{Cooper2018,Manetsch2025}.
The map $W$ is the empirical single-atom point-spread function, the difference between the mean patches of provisionally classified bright and dark shots, clipped at zero and normalized to $\max W=1$.
The data sets comprise 616 and 561 shots of the five-site array at $x=80\,\upmu$m and $1\,\upmu$m, respectively, each with ten consecutive images.
Shots in which a site exceeds the bright-peak mean by more than $4\sigma$ in the first image are attributed to residual double occupancy and discarded (4 and 0 shots observed in two positions, respectively); the effect on the fidelity is negligible.

The imaging fidelity is characterized without assumptions on the photon statistics using the three-image method of Ref.~\cite{Manetsch2025}.
The weighted counts of three consecutive images are binarized at a common threshold $\theta$, and the frequencies of the eight outcome triplets, evaluated over $5\times612=3060$ and $5\times561=2805$ three-image sequences (five sites per shot), are fitted to a four-parameter model with initial filling fraction $f$, per-image survival probability $p_{\mathrm{s}}$, and independent false-positive and false-negative probabilities $1-\mathcal{F}_0$ and $1-\mathcal{F}_1$.
The imaging fidelity is
\begin{equation} \label{eq:F_imaging}
     F_{\mathrm{img}} = f\mathcal{F}_1+(1-f)\mathcal{F}_0,
\end{equation}
and $\theta$ is chosen to maximize $F_{\mathrm{img}}$, with false-positive probabilities $1-\mathcal{F}_0=0.20\%$ and $0.16\%$, false-negative probabilities $1-\mathcal{F}_1=3\times10^{-6}$ and $1.6\times10^{-4}$, and filling $f=0.745(15)$ and $0.751(16)$, for $x= 80\,\upmu$m and 1\,$\upmu$m, respectively. 
Uncertainties in the reported imaging fidelities and fitted parameters are 95\% confidence intervals from bootstrap resampling of complete shots with $W$ and $\theta$ held fixed.
The survival in Fig.~\ref{fig:imaging}b is the fraction of sites bright in the first image that remain bright in imaging count $j$, evaluated with the same $W$ and $\theta$ over ten consecutive images.

\subsection{Cavity spectroscopy and photon counting}
\label{md:cavspec}

After transport to the side-trap configuration, the atom-cavity response is measured in reflection using a weak probe light.
During spectroscopy, the auxiliary locking light is extinguished to prevent residual off-resonant perturbations of the atoms, while the 759-nm heating power is held fixed at its mean feedback value over the preceding 20-ms interval.
For each experimental realization, a 1-ms weak-probe pulse measures the atom-cavity reflection at a selected probe detuning, with the incident power adjusted according to the atom number to suppress probe-induced optical pumping out of the clock state.
Immediately afterward, the probe frequency is swept across the cavity resonance for 20~ms at an incident power approximately 100 times higher than that used for the atom-cavity measurement.
At the beginning of this sweep, the strong 1389-nm probe rapidly depumps the atoms from the ${}^{3}\mathrm{P}_{0}$ state through excitation to ${}^{3}\mathrm{D}_{1}$, thereby removing them from resonant coupling to the cavity.
Because the depumping occurs rapidly compared with the duration of the sweep, its contribution to the recorded spectrum is negligible, and the resulting trace is treated as the empty-cavity spectrum.
This spectrum is used to extract the cavity resonance frequency and external coupling rate for each experimental realization.
The resonance frequencies extracted from these empty-cavity scans indicate that the cavity resonance remains within $\pm0.20$\,MHz of the atomic transition in 68\% of realizations over a 100-ms hold, confirming sufficient stability throughout the spectroscopy sequence. 

The reflected probe is spectrally filtered before detection to suppress Raman-scattered photons generated in the standard silica-fiber sections by the 759-nm light used for photothermal cavity frequency control.
The filtering stage comprises two volume Bragg gratings with a bandwidth of 60\,GHz, followed by a filter cavity with a full width at half maximum of 110\,MHz at a finesse of 350.
The filtered photons are detected using a superconducting nanowire single-photon detector with a quantum efficiency of $\sim 80\%$~\cite{Miki2013}, giving an overall cavity-output-to-detector efficiency of $\sim 30\%$, including filtering and optical transmission losses.
The remaining background count is dominated by room-temperature blackbody radiation coupled into the fiber rather than by leakage through the spectral filters.

\subsection{Fitting procedure for the cavity-reflection spectra}
\label{nd:fitmodel}

The cavity-reflection spectra are analyzed using a single-sided-cavity input-output model~\cite{Kikura2025caps}. 
For $N$ atoms with a common effective single-atom coupling $g$ and collective coupling $G^2=Ng^2$, the amplitude reflection coefficient is
\begin{equation}\label{eq:reflection-coef}
r_N(G, \Delta_{\mathrm{c}},\Delta_{\mathrm{a}})
=
1-
\frac{2\kappa_{\mathrm{ex}}}
{\kappa_{\mathrm{ex}}+\kappa_{\mathrm{in}}-i\Delta_{\mathrm{c}}
+G^2/(\gamma-i\Delta_{\mathrm{a}})},
\end{equation}
where $\Delta_{\mathrm{c}}$ and $\Delta_{\mathrm{a}}$ are the probe detunings from the cavity and atomic resonances, respectively, and $\gamma/2\pi=0.24\,\mathrm{MHz}$ \cite{Beloy2012} is the atomic polarization-decay rate. 

For each experimental realization $j$ with atom number $N_j$, the atomic detuning is $\Delta_{\mathrm{a},j}=2\pi(f_{\mathrm{p},j}-f_{\mathrm{a}})$, where $f_{\mathrm{p},j}$ is the programmed probe frequency, and the atomic resonance $f_{\mathrm{a}}$ defines zero detuning.
The cavity is locked to the atomic resonance, which means $\Delta_{\mathrm{c},j}=\Delta_{\mathrm{a},j}$ for all shots; 
the per-shot cavity resonance $f_{\mathrm{c},j}$ extracted from the in-shot empty-cavity sweep is used only to reject realizations for which $|f_{\mathrm{c},j}-\bar f_{\mathrm{c}}|>2\,\mathrm{MHz}$, where $\bar f_{\mathrm{c}}$ is the run median.
The internal loss rate is fixed at $\kappa_{\mathrm{in}}/2\pi=1.28(6)\,\mathrm{MHz}$, and $\kappa_{\mathrm{ex}}$ is fixed to the value obtained from a Poisson fit of the empty-cavity sweeps.
For Fig.~\ref{fig:overview}e and Fig.~\ref{fig:coupling}, the fit to the 20-ms empty cavity spectrum data aggregated over each data collection yields $\kappa_{\mathrm{ex}}/2\pi=1.21(1)\,\mathrm{MHz}$, while for the data shown in Fig.~\ref{fig:collective}, $\kappa_{\mathrm{ex}}/2\pi=1.41(1)\,\mathrm{MHz}$, due to a slightly different configuration for cavity tuning. 
For Fig.~\ref{fig:tunability}c the outcoupling rate was $\kappa_{\mathrm{ex}}/2\pi=2.33(1)\,\mathrm{MHz}$, resulting in $\kappa_\mathrm{ex} = 1.82(9) \kappa_\mathrm{in}$.

Realizations are further postselected based on the cavity lock status (no failure) and the analysis result of the fluorescence images taken before pumping the atoms to the clock state, which determines the loaded atom number.
Photons detected during the first $250\,\upmu\mathrm{s}$ of the 1-ms probe pulse are summed to give the photon count $n_j$, modeled as $n_j\sim\operatorname{Poisson}(\lambda_j)$ with
\begin{equation}
\lambda_j = C_0\,\eta(f_{\mathrm{p},j})\left|r_{N_j}(G,\Delta_{\mathrm{c},j},\Delta_{\mathrm{a},j})\right|^2+b,
\end{equation}
where $C_0$ is the count amplitude, $b$ the additive background, and $\eta(f_{\mathrm{p}})$ the independently calibrated Gaussian diffraction efficiency of the probe acousto-optic modulator ($1/e$ half-width $18.5\,\mathrm{MHz}$; $\eta=0.81$--$0.85$ at $\pm8\,\mathrm{MHz}$), which scales the reflected signal but not the background.
The negative log-likelihood $\sum_j[\lambda_j-n_j\ln\lambda_j]$ is minimized over $\{g,C_0,b\}$ for the single-atom data and over $\{G,C_0,b\}$ for the collective data, and the $1\sigma$ uncertainty in $g$ is the inverse square root of the curvature of the profile negative log-likelihood, with $C_0$ and $b$ re-optimized at each value of $g$.
Where the fit converges to $g=0$ (near the nodes in Fig.~\ref{fig:coupling}c), this curvature vanishes because the model depends on $g$ only through $g^2$, and we instead report the upper edge of the 95\% profile-likelihood confidence interval as an upper bound; these points are drawn at $g=0$ with an error bar extending to the bound.

\subsection{Modeling the finite-temperature and clock-state-depumping effect on cavity reflection spectroscopy}

\subsubsection{Finite temperature effect}
\label{md:thermal}

The position-dependent single-atom coupling is $g(x,y,z)=g(x,z)\,|\cos\beta y|$, with $x$ as the radial distance from the nanofiber axis, $y$ as the position along the nanofiber, $z$ as the position along the tweezer axis, and $\beta=2\pi n_\mathrm{eff}/\lambda$. 
The coupling $g(x,z)$ is calculated from the $\mathrm{HE}_{11}$ mode of the 550-nm nanofiber for the $\pi$-polarized ${}^{3}\mathrm{P}_{0}\rightarrow{}^{3}\mathrm{D}_{1}$ transition and the cavity mode volume, giving $g_0/2\pi=g(1\,\upmu\mathrm{m},0)/2\pi=2.95\,\mathrm{MHz}$ at the trap minimum.
A thermal atom with a finite spatial distribution has a root-mean-squared coupling $\bar g=\langle g^2\rangle^{1/2}$ over its position distribution in the trap, with widths $\sigma_i=\sqrt{k_BT/m\omega_i^2}$ for trap frequencies $\omega_{x,y}/2\pi=21\,\mathrm{kHz}$ and $\omega_z/2\pi=3.7\,\mathrm{kHz}$.
The three-dimensional average is evaluated numerically at $T=2.5(3)\,\upmu\mathrm{K}$ ($\sigma_{x,y}=84\,\mathrm{nm}$, $\sigma_z=0.47\,\upmu\mathrm{m}$), resulting in $q_T=\bar g/g_0=0.83(2)$, dominated by the spread along the tweezer axis; the spread along $y$ reduces the standing-wave contrast in Fig.~\ref{fig:coupling}c.

\subsubsection{Clock-state preparation and depumping}
\label{md:depumping}

The probe duration window of $250\,\upmu\mathrm{s}$ was chosen to reduce the effect of depumping from the clock state.
We model the depumping as a decay of the clock-state population at a rate $\Gamma\,s(\Delta)$, where $\Gamma$ is the resonant depumping rate and $s(\Delta)=S(\Delta)/S(0)$ is the free-space scattering probability from the atom per incident probe photon~\cite{Welte2018,Reiserer2015},
\begin{equation}
S(\Delta)=\frac{4\kappa_{\mathrm{ex}}\gamma\,\bar g^2}{|\gamma-i\Delta|^2\,\left|\kappa_\mathrm{ex}+\kappa_\mathrm{in}-i\Delta+\bar g^2/(\gamma-i\Delta)\right|^2},
\end{equation}
normalized to its resonant value.
A joint fit of the single-atom spectra (Fig.~\ref{fig:overview}e) in ten successive 100-$\upmu$s bins of the 1-ms probe, with a fixed $g$ and a coupled fraction decaying as $e^{-\Gamma s(\Delta)t}$, yields $\Gamma=2.5(6)\,\mathrm{ms}^{-1}$; the final bins show a nearly empty cavity spectrum.
The expected cavity reflection spectrum for a single atom is then the mixture $R'(\Delta_j) = (1-p(\Delta_j))\,|r_0(0,\Delta_j)|^2+p(\Delta_j)\,|r_1(\bar g,\Delta_j)|^2$
with $\Delta_j=\Delta_{\mathrm{a},j}=\Delta_{\mathrm{c},j}$ and $p(\Delta)=p_\mathrm{pump}\,\bar p(\Delta)$ as the coupled weight, where $p_\mathrm{pump}=0.905(35)$ is the clock-state preparation efficiency (since the presence of atoms is determined in ${}^{1}\mathrm{S}_{0}$ before pumping, a fraction $1-p_\mathrm{pump}$ of the realizations labeled $N=1$ contains an uncoupled atom; the value is taken from the bounds reported in Methods Sec.~\ref{md:atomprep}) and $\bar p(\Delta)=[1-e^{-\Gamma s(\Delta)\tau_\mathrm{w}}]/[\Gamma s(\Delta)\tau_\mathrm{w}]$ is the clock-state population averaged over the window $\tau_\mathrm{w}=250\,\upmu\mathrm{s}$; on resonance $p(0)=0.67$.
Numerically sampling the mixture model and following the same fitting procedure as in the experiment (see previous section), returns $g=q_\mathrm{pump}\,\bar g$ with $q_\mathrm{pump}=0.86(5)$.
The predicted coupling at the side-trap position, $g_0 q_T q_\mathrm{pump}/2\pi=2.11(11)\,\mathrm{MHz}$, is $1.2\sigma$ above the measured $1.86(17)\,\mathrm{MHz}$, leaving a residual factor $q_\mathrm{res}=0.88(9)$.
We attribute $q_\mathrm{res}$ to a higher temperature along the weakly confined tweezer axis than that measured by release-and-recapture thermometry, which is mainly sensitive to radial motion, as well as to residual cavity drift.
These technical limitations can be mitigated by improved optical pumping, additional repumping lasers during the spectroscopy, and stronger confinement or more efficient atom cooling.

For the position scans of Figs.~\ref{fig:coupling}b,c, the model is $q'\,\bar g(x_0,y_0;T)$ with $T=2.5(3)\,\upmu\mathrm{K}$ fixed and $q'=q_\mathrm{pump}'q_\mathrm{res}=0.69(9)$, where $q_\mathrm{pump}'=0.79(6)$ is re-evaluated since $\Gamma$ scales with the resonant scattering rate and is averaged over the scanned positions.
The only fitted quantity is the origin of $y_0$, and the shaded band is the envelope of the model over $T\pm0.3\,\upmu\mathrm{K}$, with $q_\mathrm{res}$ re-derived at each $T$, and $q'\pm\delta q'$.

\subsection{Nanofiber diameter design for QPU integration}

For integrating a nanofiber cavity with a reconfigurable atom array, it is desirable to choose the nanofiber diameter so as to suppress backreflection of the tweezer light during atom transport while maintaining strong guided-mode coupling at the target atomic transition.
We first analyze the diameter dependence of tweezer-light scattering from a nanofiber and then combine the resulting backscattering minima with the coupling requirement to identify suitable combinations of nanofiber diameter and tweezer wavelength.

\subsubsection{Plane-wave scattering by a nanofiber}

The plane-wave scattering from an infinitely long dielectric cylinder \cite{wait1955}, known as Lorenz--Mie scattering by a cylinder, provides a baseline model for the nanofiber-diameter-dependent reflectivity. Since the system is translationally invariant along the nanofiber axis $y$, the scattering problem is described in the transverse cylindrical coordinates $(\rho,\phi)$. For a normally incident, linearly polarized plane wave with vacuum wavelength $\lambda_\text{tweezer}$ and wavenumber $k=2\pi/\lambda_\text{tweezer}$, assuming vacuum outside the nanofiber and refractive index $n$ for fused silica, the scattered-field amplitude normalized to the incident plane-wave amplitude can be expanded as
\begin{align}
    \psi(k\rho,\phi)
    =
    \sum_{\ell=-\infty}^{+\infty}
    i^\ell c_\ell H_\ell^{(1)}(k\rho)e^{i\ell\phi},
\end{align}
where $H_\ell^{(1)}$ is the Hankel function of the first kind. 
At normal incidence, the two linear-polarization channels decouple, and the complex partial-wave coefficients for electric fields parallel and perpendicular to the nanofiber axis are
\begin{align}
    c_\ell^{\parallel}
    &=
    \frac{n J'_\ell(nka)J_\ell(ka)-J_\ell(nka)J'_\ell(ka)}
    {H_\ell^{(1)\prime}(ka)J_\ell(nka)-nJ'_\ell(nka)H_\ell^{(1)}(ka)},\\
    c_\ell^{\perp}
    &=
    \frac{J'_\ell(nka)J_\ell(ka)-nJ_\ell(nka)J'_\ell(ka)}
    {nH_\ell^{(1)\prime}(ka)J_\ell(nka)-J'_\ell(nka)H_\ell^{(1)}(ka)},
\end{align}
where $J_\ell$ is the Bessel function of the first kind, the prime denotes differentiation with respect to the argument, and $a=d/2$ is the nanofiber radius, such that $ka=\pi d/\lambda_\text{tweezer}$.
The normalized backward-scattered intensity $|\psi(k\rho,\pi)|^2$ exhibits minima at particular values of $ka$ due to destructive interference among the complex partial-wave amplitudes. 

For the perpendicular-polarization channel, using the refractive index of fused silica at $\lambda_\text{tweezer}=759\,\mathrm{nm}$, the far-field ($k\rho\gg1$) minimum in the experimentally relevant diameter range occurs at $ka\simeq2.48$, corresponding to $d/\lambda_\text{tweezer}\simeq0.790$ and hence $d\simeq599\,\mathrm{nm}$.
At finite distances, the minimum position depends on $k\rho$ through the Hankel functions but remains close to its far-field value. 
For this particular minimum, even at the nanofiber surface, $k(\rho-a)=0$, the minimum occurs at $ka\simeq2.46$, or $d\simeq594\,\mathrm{nm}$, corresponding to a shift of only about $0.9\%$ from its far-field value.

\subsubsection{Focused-beam scattering and model comparison}

While this plane-wave model provides a useful estimate of the nanofiber diameter at which backward scattering is suppressed, the depth and angular structure of the suppression for a focused tweezer beam depend on the spatial mode of the incident field.
In particular, the angular components of a focused beam excite the cylindrical partial waves with different amplitudes and phases, and their coherent interference must be retained when evaluating the backward-scattered field.

We therefore compare the plane-wave cylinder model with a Gaussian-beam treatment based on generalized Lorenz--Mie theory (GLMT)~\cite{ren1997} and with three-dimensional finite-difference time-domain (FDTD) simulations.
As shown in Fig.~\ref{fig:reflection-model-comparison}, all three approaches yield similar diameter-dependent locations of the principal backward-scattering minima and broadly similar suppression patterns, with quantitative differences in the depth and angular structure of the minima.
This agreement indicates that the plane-wave model captures the relevant diameter range and the overall suppression behavior, while the focused-beam calculations account for quantitative modifications arising from the finite spatial mode of the optical tweezer.
We thus use the plane-wave model as a simple guide to identify combinations of nanofiber diameter and tweezer wavelength that suppress backward scattering. 

\subsubsection{Diameter--wavelength design across atomic species}

To extend this design principle across atomic species, we combine the backscattering minima predicted by the dielectric-cylinder model with the nanofiber-diameter requirements for strong guided-mode coupling at the relevant atomic transitions.
We calculate successive minima of the backward-scattered field as a function of the tweezer wavelength for polarization perpendicular to the nanofiber axis. 
For an approximate estimate of the diameter range compatible with strong atom-cavity coupling, we use the geometric condition \(d/\lambda_{\mathrm{a}}\lesssim0.6\), where \(\lambda_{\mathrm{a}}\) is the resonant atomic-transition wavelength, with the precise limit depending on the atom-surface separation.
The overlap between the resulting diameter ranges and the low-backreflection regions identifies candidate combinations of tweezer wavelength and nanofiber diameter for low-perturbation QPU integration (Fig.~\ref{fig:diameter-design}).

For ${}^{171}\mathrm{Yb}$, the diameter range compatible with the telecom-band transitions at $1389$--$1539\,\mathrm{nm}$ overlaps the relevant backreflection minima near $d=599\,\mathrm{nm}$ for a $759$-nm tweezer and $d=536\,\mathrm{nm}$ for a $486$-nm tweezer. 
The same construction identifies compatible design windows for 
other species previously used in atom-nanofiber interfaces, including ${}^{87}\mathrm{Rb}$, ${}^{133}\mathrm{Cs}$ and ${}^{88}\mathrm{Sr}$ at the representative tweezer wavelengths shown in Fig.~\ref{fig:diameter-design}~\cite{Vetsch2010,Kato2015,Kestler2023}. 
These results show that the nanofiber diameter and tweezer wavelength can be jointly selected to suppress tweezer-induced perturbations while retaining coupling to the target atomic transition.

\subsection{Nanofiber diameter uniformity and reproducibility}\label{md:uniformity}

To quantify millimeter-scale diameter uniformity and fabrication reproducibility, we fabricated two sets of nanofibers with waist diameters of 460 and 650\,nm, using fixed fabrication conditions for each set, and measured their diameter profiles \(D(y)\) over the central 1-mm region by scanning electron microscope (Fig.~\ref{fig:SEM}a,b). 
The distribution of the mean waist diameter $D_m$ across each set characterizes fabrication reproducibility, whereas the intra-device standard deviation $\sigma_D$ quantifies diameter uniformity along each nanofiber. 
The two sets yielded mean values of $D_m=459(5)\,\mathrm{nm}$ and $D_m=651 (8)\,\mathrm{nm}$ for \(N>10\) devices per set, respectively, where the uncertainties denote the standard deviations across devices. 
All measured nanofibers exhibited below-percent level uniformity over 1\,mm, \(\sigma_D/D_m<1\%\) (Fig.~\ref{fig:SEM}c,d).

The 460 and 650-nm fabrication sets target minima in tweezer-light backreflection at representative \(^{87}\mathrm{Rb}\) tweezer wavelengths of 1013 and 821\,nm, respectively. 
The corresponding diameter ranges are compatible with \(^{87}\mathrm{Rb}\) telecom-band transitions, while the 460-nm range also supports coupling on the 780-nm D\({}_2\) transition. 
These results demonstrate reproducible fabrication within species-specific low-reflection diameter windows while maintaining sub-percent diameter variation along each nanofiber.

\subsection{Time- and channel-multiplexed remote entanglement generation}
The proposed quantum networking architecture (Fig.~\ref{fig:transport}) uses a networking layer where atoms are coupled to the cavity mode in a side-trap configuration, and a displaced transport layer where atom arrays can be moved across the nanofiber with negligible perturbation. Transfer between the two layers is performed by three-dimensional transport.
In addition to a simplified time-multiplexed remote entanglement generation scheme shown in Fig.~\ref{fig:transport}d, we propose a more efficient pipelined operating sequence for time- and channel-multiplexed remote entanglement generation (Fig.~\ref{fig:timemux}).
Two atom arrays are positioned on opposite sides of the nanofiber in the networking layer, with one coupled to the cavity and the other held at a decoupled position 5\,$\upmu$m away from the nanofiber axis.
During a networking window (ii, iv), atoms in the coupled array are sequentially addressed for time-multiplexed entanglement attempts, while the previously coupled array is transferred to the transport layer and replaced by a new array at the decoupled position. 
The arrays are then switched (i, iii), so that a new set of atoms can be coupled for entanglement generation attempts. 
This overlap of networking with array replacement amortizes preparation and transport overheads and sustains a high networking duty cycle, with only the time overhead relevant to the entanglement generation rate being the array switching. 
The same sequence can be operated in parallel across multiple nanofiber cavities for channel multiplexing.

To estimate concrete time overheads for operations in Fig.~\ref{fig:timemux}, we assume a tweezer with radial and axial trap frequencies of $2\pi\times100$\,kHz and $2\pi\times21$\,kHz operating shortcut-to-adiabaticity transport~\cite{Hwang25} over 5\,$\upmu$m; the in-plane switch of the coupled array over 5\,$\upmu$m takes less than 40\,$\upmu$s, and the transfer between the transport and networking layers along the tweezer axis takes 130\,$\upmu$s, for up to $\Delta n=1$ motional excitation, where the transfer between the two layers runs in parallel with the remote entanglement generation.
The rate quoted in the main text follows the sequential cavity-assisted photon scattering protocol and multiplexed rate model of Ref.~\cite{Kikura2025caps}, with internal cooperativity at $C_\mathrm{in}=5.6$, $\gamma/2\pi=0.24$\,MHz, and $\kappa_\mathrm{ex}=\sqrt{1+2C_\mathrm{in}}\,\kappa_\mathrm{in}$. 
50 atoms sequentially attempt entanglement once each without in-cavity reset, and a shuttling overhead of 40\,$\upmu$s is incurred.
A heralded infidelity of $10^{-2}$ requires a photon width of $\sigma_t\simeq1/\gamma=0.68\,\upmu$s at $C_\mathrm{in}=5.6$~\cite{Kikura2025caps}, for which the success probability per attempt is $19\,\%$ and the time-multiplexed rate is 40\,kHz for a single channel.
For this estimation, as in Ref.~\cite{Kikura2025caps}, we assume the single photon is supplied from the same atom-cavity system with a success probability determined by the $C_\mathrm{in}$ value used here.
Single-photon detectors, fibers, and other optical losses are not included and reduce the rate proportionally, while channel multiplexing increases the rate proportionally.

\clearpage

\setcounter{equation}{0}
\renewcommand{\theequation}{S\arabic{equation}}
\renewcommand{\theHequation}{S\arabic{equation}}
\setcounter{figure}{0}
\renewcommand{\thefigure}{S\arabic{figure}}
\renewcommand{\theHfigure}{S\arabic{figure}}
\newcounter{supplsec}
\renewcommand{\thesupplsec}{S\arabic{supplsec}}
\renewcommand{\theHsupplsec}{S\arabic{supplsec}}

\newcommand{\supsection}[1]{
  \refstepcounter{supplsec}
  \begin{center}
  \noindent\textbf{\thesupplsec: #1}\par
  \end{center}
}
\newcounter{supplsubsec}[supplsec]
\renewcommand{\thesupplsubsec}{\thesupplsec.\arabic{supplsubsec}}
\newcommand{\supsubsection}[1]{
  \refstepcounter{supplsubsec}
  \begin{center}
  \noindent\textbf{\thesupplsubsec\quad #1}\par
  \end{center}
}

\begin{figure*}[h]
    \centering
    \includegraphics[width=\textwidth]{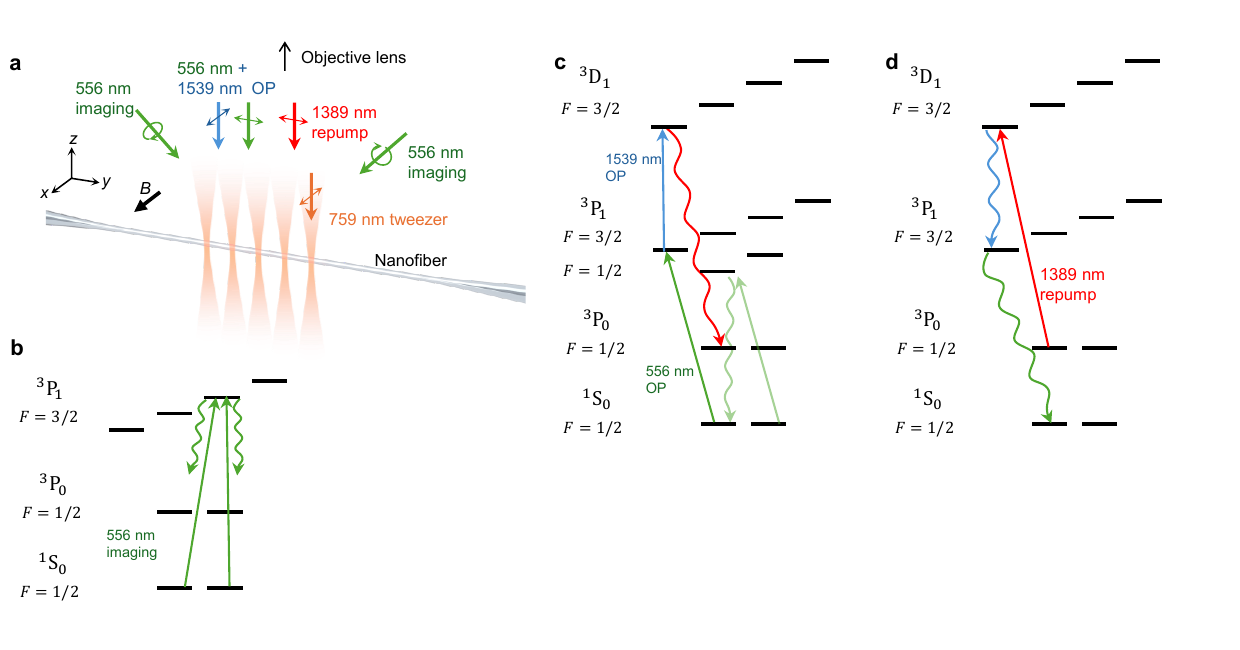}
    \caption{\textbf{${}^{171}\mathrm{Yb}$ level diagrams and laser beam geometry.}
    \textbf{a},
    Geometry of the tweezer, optical-pumping (OP), depumping and imaging beams relative to the nanofiber and imaging axes, with a 20~Gauss bias magnetic field defining the quantization axis.
    \textbf{b}, Fluorescence imaging on the \(^{1}\mathrm{S}_{0},\,F=1/2 \leftrightarrow {}^{3}\mathrm{P}_{1},\,F=3/2\) transition at \(556\,\mathrm{nm}\).
    \textbf{c}, Optical pumping into the \(^{3}\mathrm{P}_{0},\,F=1/2,\,m_F=-1/2\) state using simultaneous excitation at \(556\,\mathrm{nm}\) (\(^{1}\mathrm{S}_{0}\rightarrow{}^{3}\mathrm{P}_{1}\)) and \(1539\,\mathrm{nm}\) (\(^{3}\mathrm{P}_{1}\rightarrow{}^{3}\mathrm{D}_{1}\)).
    \textbf{d}, Depumping to \(^{1}\mathrm{S}_{0}\) state via excitation on the \(1389\,\mathrm{nm}\) (\(^{3}\mathrm{P}_{0}\rightarrow{}^{3}\mathrm{D}_{1}\)) transition after cavity spectroscopy, returning the atoms to the fluorescence-imaging cycle.
    \label{fig:geometry}}
\end{figure*}

\begin{figure*}[ht]
    \centering
    \includegraphics[width=\textwidth]{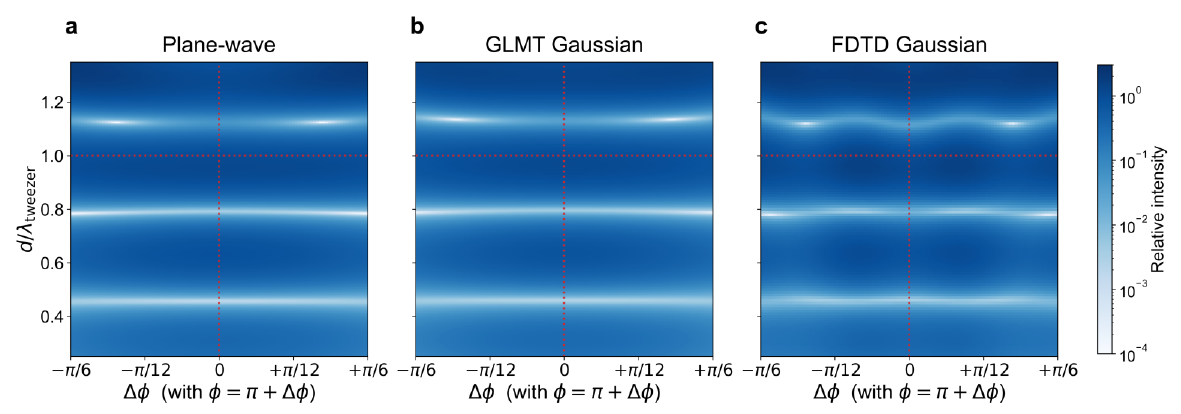}
    \caption{\textbf{Backward-scattering suppression for plane-wave and focused-Gaussian illumination of nanofibers.}
    Relative co-polarized backward-scattered intensity as a function of the normalized nanofiber diameter $d/\lambda_\text{tweezer}$ and angular deviation $\Delta\phi$ from the backward direction, $\phi=\pi+\Delta\phi$.
    \textbf{a}, Plane-wave cylinder model.
    \textbf{b}, Gaussian-beam calculation based on GLMT~\cite{ren1997}.
    \textbf{c}, Three-dimensional FDTD simulation of a focused Gaussian beam.
    The incident field is linearly polarized perpendicular to the nanofiber axis, with a nanofiber refractive index of $n=1.45$ and a Gaussian-beam waist of $w_0/\lambda_\text{tweezer}=1.2$ for the Gaussian-beam calculations.
    The intensity is sampled on a plane perpendicular to the backward direction, with its on-axis distance from the nanofiber axis set to $k\rho_0=2\pi\times4.5/0.759$. For $\lambda_\text{tweezer}=759\,\mathrm{nm}$, this corresponds to $\rho_0=4.5\,\upmu\mathrm{m}$, consistent with the spatial geometry used in Fig.~\ref{fig:transport}b.
    In each panel, the intensity is normalized to its value at $d/\lambda_\text{tweezer}=1$ and $\Delta\phi=0$; the dotted lines indicate these reference values.
    The principal suppression minima occur at similar values of $d/\lambda_\text{tweezer}$ in all three models, whereas their depth and angular structure depend on the treatment of the incident spatial mode.
    \label{fig:reflection-model-comparison}}
\end{figure*}

\begin{figure*}[ht]
    \centering
    \includegraphics[width=0.85\textwidth]{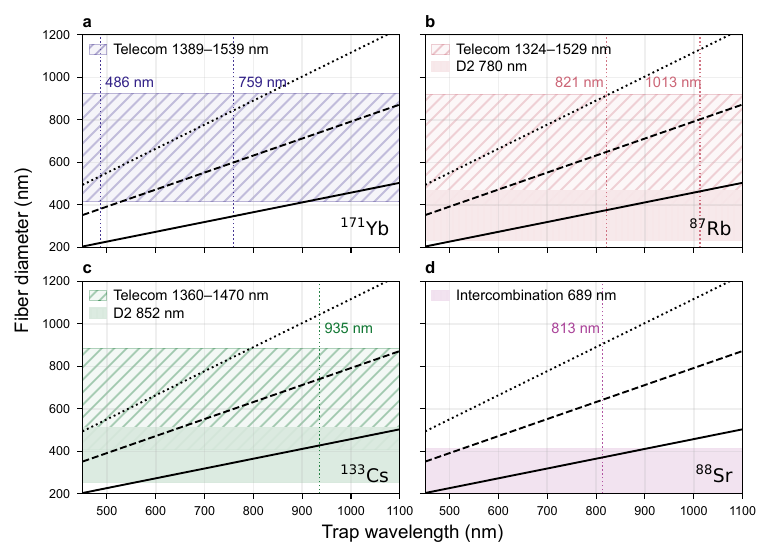}
    \caption{\textbf{Nanofiber-diameter design for lossless tweezer transport for various atomic species used for neutral-atom quantum computing.}
    Candidate nanofiber diameters are shown as a function of tweezer wavelength for ${}^{171}\mathrm{Yb}$, ${}^{87}\mathrm{Rb}$, ${}^{133}\mathrm{Cs}$, and ${}^{88}\mathrm{Sr}$.
    The solid, dashed, and dotted black curves mark successive minima of the calculated far-field backscattering for tweezer light polarized perpendicular to the nanofiber axis, in order of increasing fiber diameter.
    Vertical dotted lines mark representative tweezer wavelengths, and the colored horizontal bands show usable nanofiber-diameter ranges for the labeled atom-cavity transitions, with hatched bands denoting telecom-band transitions and solid bands denoting the D2 or intercombination transitions.
    Their intersections identify diameter and wavelength combinations that suppress tweezer backreflection while supporting coupling to the desired atomic transition.
    For ${}^{171}\mathrm{Yb}$ at a tweezer wavelength of 759~nm (486~nm), the relevant backscattering minimum occurs near a fiber diameter of 599~nm (536~nm).}
    \label{fig:diameter-design}
\end{figure*}

\begin{figure*}[ht]
    \centering
    \includegraphics[width=0.99\linewidth]{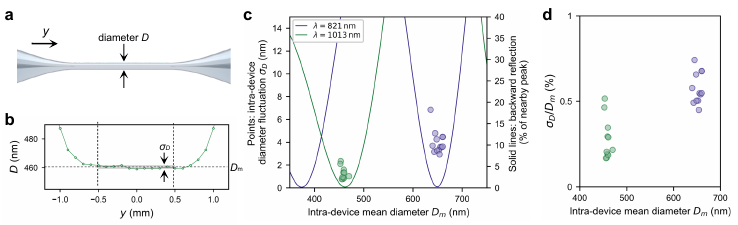}
    \caption{\textbf{Nanofiber diameter uniformity and fabrication reproducibility.}
    \textbf{a}, Schematic of the nanofiber profile, showing the diameter $D(y)$ along the fiber axis $y$.
    \textbf{b}, Representative measured diameter profile. 
    The mean waist diameter $D_m$ and its standard deviation $\sigma_D$ are evaluated over the central $1$-mm region indicated by the vertical dashed lines.
    \textbf{c}, Intra-device diameter variation $\sigma_D$ versus mean waist diameter $D_m$ for nanofibers with target diameters of 460 and 650\,nm. Each point represents an independently fabricated device, showing the device-to-device distribution within each fabrication set.
    Solid curves show the calculated backward reflection at representative \(^{87}\mathrm{Rb}\) tweezer wavelengths of \(\lambda=821\) and \(1013\,\mathrm{nm}\), for which the corresponding nanofiber diameters are compatible with telecom-band transitions. 
    Each curve is normalized to its nearest local maximum.
    \textbf{d}, Relative diameter variation, \(\sigma_D/D_m\), which remains below 1\% for all measured devices.
    \label{fig:SEM}}
\end{figure*}

\begin{figure*}[ht]
    \centering
    \includegraphics[width=1\textwidth]{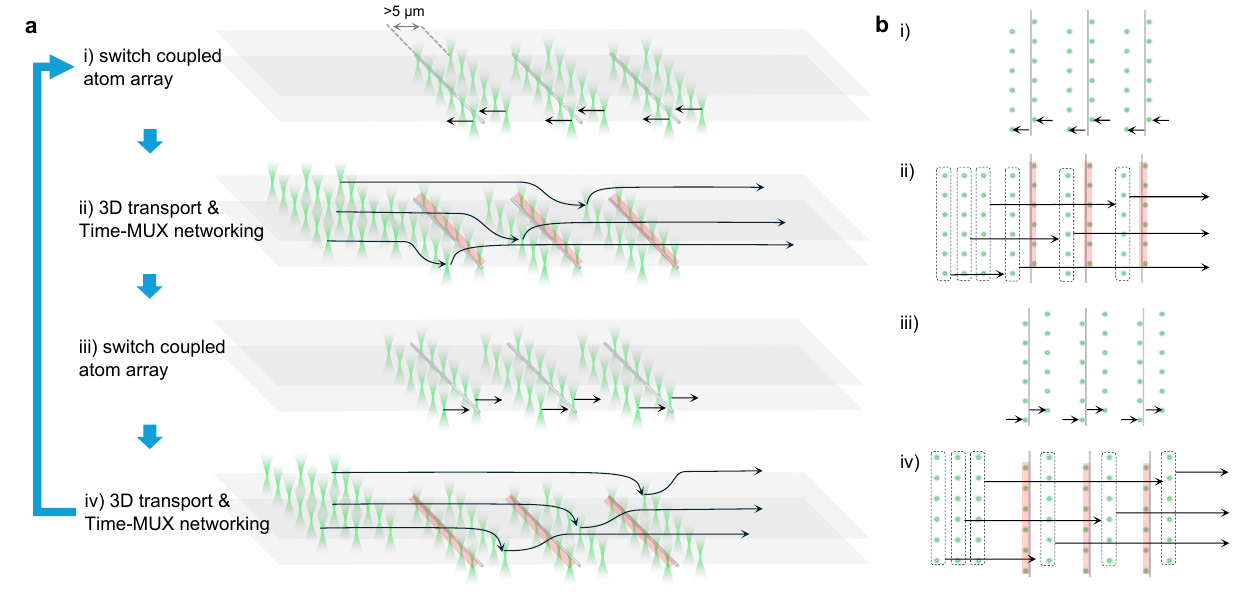}
    \caption{\textbf{Time- and channel-multiplexed remote entanglement generation.}
    \textbf{a}, Lossless transport across the nanofiber demonstrated in Fig.~\ref{fig:transport} can be used for time- and channel-multiplexed remote entanglement generation.
    This follows a pipelined operating sequence, where two atom arrays are positioned on opposite sides of the nanofiber in the networking layer, allowing one array to undergo time-multiplexed remote entanglement generation while the other is transported out and replenished through the transport layer.
    The continuous operation cycles through four steps; 
    (i) The coupled atom array is switched in the transport layer and replaced by a new array at the decoupled position, in approximately $40~\upmu\mathrm{s}$; here, thanks to a very small mode volume of the nanofiber cavity, the switch requires a short transport length. 
    We assume 5 $\upmu$m away from nanofiber axis, which makes the atom-cavity coupling two orders of magnitude smaller than at the side-trap positions.
    (ii) Time-multiplexed networking proceeds with atoms coupled to the cavity for entanglement attempts (red shading), while the remaining atom arrays are shuttled to bring the new set of atom array near the nanofiber while moving the post-entanglement attempt atoms to the processing region, such as for entanglement distillation.
    (iii) The coupled atom array is switched again.
    (iv) Time-multiplexed networking proceeds again while the remaining atom arrays are shuttled, and this cycle repeats.
    \textbf{b}, top view of the same operating sequence in \textbf{a}.
    \label{fig:timemux}}
\end{figure*}

\end{document}